\documentclass[prb,12pt,onecolumn,superscriptaddress]{revtex4-1} 
\usepackage{amssymb,graphicx,color,amsmath,xfrac,bm,url,oldstyle,soul}
\usepackage[mathletters]{ucs}
\usepackage[utf8x]{inputenc}

\definecolor{lightblue}{rgb}{0.17,0.39,1}
\definecolor{lightgreen}{rgb}{0.67,0.81,0.08}
\definecolor{lightred}{rgb}{1,0.05,0.52}

\newcommand{\curly}[1]{\left\{#1\right\}}

\newcommand{\hide}[1]{{}}							
\newcommand{\te}[1]{{\text{#1}}}								
						
\newcommand{\B}{\text{B}} 
\renewcommand{\H}{\text{H}}
\newcommand{\LSCO}{$\te{La}_{2-x}\te{Sr}_x\te{CuO}_4$\ }

\begin{document}
\title{Energy-scale competition in the Hall resistivity of a strange metal}

\author{A.~Shekhter}  
\email[Email: ]{arkadyshekhter@gmail.com}
\affiliation{Los Alamos National Laboratory, Los Alamos, NM 87545, USA}
\author{K.~A.~Modic} 
\affiliation{Institute of Science and Technology Austria, Am Campus 1, 3400 Klosterneuburg, Austria} 
\author{L.~E.~Winter}
\author{Y.~Lai}
\author{M.-K.~Chan} 
\author{F.~F.~Balakirev} 
\author{J.~B.~Betts}
\affiliation{Los Alamos National Laboratory, Los Alamos, NM 87545, USA}
\author{S.~Komiya} 
\author{S.~Ono}
\affiliation{Central Research Institute of Electric Power Industry (CRIEPI) Materials Science Research Laboratory, 2-6-1 Nagasaka, Yokosuka, Kanagawa, Japan}
\author{G.~S.~Boebinger}
\affiliation{National High Magnetic Field Laboratory, Florida State University, Tallahassee, FL 32310, USA}
\affiliation{Department of Physics, Florida State University, Tallahassee, FL 32310, USA}
\author{B.~J.~Ramshaw}
\affiliation{Laboratory of Atomic and Solid State Physics, Cornell University, Ithaca, NY 14853, USA}
\author{R.~D.~McDonald}
\affiliation{Los Alamos National Laboratory, Los Alamos, NM 87545, USA}

\linespread{1.2} 
\begin{abstract} {\bf

Anomalous transport behavior—both longitudinal and Hall—is the defining characteristic of the strange-metal state of High-Tc cuprates. The temperature, frequency, and magnetic field dependence of the resistivity is understood within strange metal phenomenology as resulting from energy-scale competition to set the inelastic relaxation rate. The anomalously strong temperature dependence of the Hall coefficient, however, is at odds with this phenomenology. Here we report measurements of the Hall resistivity in the strange metal state of cuprates over a broad range of magnetic fields and temperatures. The observed field and temperature dependent Hall resistivity at very high magnetic fields reveals a distinct high-field regime which is controlled by energy-scale competition. This extends the strange metal phenomenology in the cuprates to include the Hall resistivity and suggests, in particular, that the direct effect of magnetic field on the relaxation dynamics of quantum fluctuations may be at least partially responsible for the anomalous Hall resistivity of the strange metal state.

}
\end{abstract} 

\date{\today}
	\maketitle	
\setlength{\parindent}{0.5cm}
\setlength{\parskip}{0.5cm}
\linespread{1.2}

The strange metal state of the high-temperature superconducting (high-Tc) cuprates is the best example of a non-Fermi liquid metal because it shows linear-in-temperature ($T-$linear) resistivity at all temperatures (Figure 1).\cite{Stormer1988, Martin1990} This $T-$linear form of the resistivity indicates a vanishing intrinsic energ­y scale—the resistivity is “scale invariant”—and is the defining characteristic of metallic quantum criticality. Scale invariance is at the core of several of the proposed microscopic mechanisms of the origin of the strange metal state. \cite{Varma1989, Anderson1992, Zaanen2004, Zaanen2019, Keimer2014} When subjected to external perturbations such as temperature, magnetic fields, or THz, infrared, and optical photons, metals characterized by a vanishing intrinsic energy scale exhibit energy scale competition -- external parameters compete to set the low-energy (infrared) physics. In the cuprates, such competition of energy scales is manifest in the competition between temperature and frequency to set the inelastic relaxation rate, $ ℏ / τ \sim  \text{max} \curly{ k_{\,\B} T , ℏ ω } $, in the optical conductivity,\cite{Basov2005}  ARPES spectra,\cite{Armitage2018} and Raman scattering.\cite{Slakey1991, Dessau2019}
 
More recently, it was shown that the strange metal state of cuprates also exhibits competition between energy scales associated with temperature and magnetic field. In transport measurements, this competition manifests itself as $B-$linear magnetoresistance at high magnetic fields and low temperatures, crossing over to $B^2$ magnetoresistance at lower magnetic fields when the thermal energy scale is higher than the energy scale set by the field.\cite{Giraldo-Gallo2018} Again, this behavior is the result of competition between the magnetic field and the temperature to set the inelastic relaxation rate, $ ℏ / τ \sim  \text{max} \curly{ k_{\,\B} T , μ_{\,\B} B } $. Together, these measurements form the basis of the strange metal phenomenology in the cuprates, where dependence on external parameters is determined by their competition to set the infrared cutoff of scale-invariant relaxation dynamics.
 
The scale-invariant behavior of longitudinal resistivity can be contrasted with the observed behavior of the Hall resistivity. At very high temperatures, the Hall resistivity in the strange metal state of cuprates has a strikingly simple form: it is proportional to magnetic field and inversely proportional to temperature, $ρ_{xy}  \propto ( μ_{\,\B}B / k_{\,\B} T ) $. \cite{Chien1991,Badoux2016} 
Unlike the longitudinal resistivity, $ρ_{xx}$, which is scale-invariant in its temperature, field, and frequency response, this $B / T$  form of the Hall resistivity is manifestly {\it non}-scale invariant -- the Hall resistivity requires a non-vanishing, intrinsic energy scale of approximately $Λ =10 K $ in the low-field regime to fix the magnitude of the observed Hall resistivity, i.e.  $ρ_{xy} \propto Λ ( μ_{\,\B}B / k_{\,\B} T ) $.
\footnote{ The high-temperature behavior of the longitudinal resistivity near the critical doping, $ρ_{xx} =  α^* T$ sets the `energy units' for resistivity via $α^* = 1.05 \mu\Omega cm / K$. The observed high-temperature behavior of the Hall resistivity in the temperature units, $ρ_{xy} / α^*$, has the form  $ρ_{xy} / α^*  = Λ ( μ_{\,\B}B / k_{\,\B} T ) $ where $Λ$ is fixed by experiment to $Λ \approx 10 K$ for chemical compositions near critical doping.\cite{Chien1991,Badoux2016} } 
Whereas the energy scale competition in the relaxation rate as observed in longitudinal transport can be attributed to the dynamics of quantum criticality in the cuprates, such attribution is not straightforward with the Hall resistivity, which in conventional metals requires a magnetic field interacting with well-defined quasiparticles. The question of whether---and how---the Hall resistivity fits within the strange metal phenomenology therefore remains unanswered. \cite{Anderson1991, Varma2001, Keimer2014, Donos2015,Ramshaw2020} 
 
Thus far, transport measurements in high magnetic fields\cite{Balakirev2009,Badoux2016, Putzke2019, Ayres2021} have not provided a clear indication of energy scale competition in the Hall resistivity of the cuprates, in part because of their liberal use of Fermi liquid phenomenology in the analysis of their measurements. What is missing is a detailed look at the temperature and magnetic field dependence of the Hall resistivity in the strange metal state at very high magnetic fields, crossing the high-field boundary where $μ_{\,\B}B > k_{\,\B} T $. 
 
To address this question experimentally, we measure the magnetic field and temperature dependence of the Hall resistivity at high magnetic fields in single crystal \LSCO close to the critical doping, $ x = $ 0.20 , $T_c =$ 34 K. \cite{ Liang2006} The sample was polished into a thin plate parallel to the copper-oxide (ab) plane and cut into a Hall bar using focused ion beam lithography (Figure 1B). This sample exhibits $T-$linear resistivity from room temperature down to the superconducting transition (Figure 1A) and linear magnetoresistance in high magnetic fields (Figure 1C) consistent with our prior magnetoresistance measurements in thin-film cuprates at a similar doping. \cite{Giraldo-Gallo2018} Figure 2A shows the Hall resistivity $ρ_{xy}$, antisymmetrized for two opposite directions of magnetic field and measured simultaneously with $ρ_{xx}$ at each temperature. 
 
The magnetic field dependence of the Hall resistivity shows a striking difference between high and low temperatures: at high temperatures, the Hall resistivity is linear in magnetic field over the entire measured field range. The Hall coefficient, $R_{\,\H} = ρ_{xy} / B $, decreases monotonically with temperature above 100 K (Figure 2A), consistent with its inverse-T dependence at very high temperatures.\cite{Chien1991, Badoux2016} At intermediate temperatures, below 100 K and above 15 K, a new phenomenon sets in: the observed Hall resistivity is no longer linear in magnetic field over the entire measured field range. This manifests as a Hall coefficient that decreases monotonically with magnetic field (Figure 2C).
The disparity of the magnetic field dependence of the Hall resistivity in these two regimes is made more evident by considering the temperature and magnetic field dependence of the ‘differential Hall coefficient’, $r_{\,\H} ≡ d ρ_{xy} / d B $.  Figure 3A shows the temperature dependence of the differential Hall coefficient at low field (5 T, in blue) and high field (60 T, in red). The two are nearly coincident for  temperatures above 100 K but diverge markedly at lower temperatures. 
The low-field differential Hall coefficient depends weakly on temperature below 100 K until it is truncated by superconductivity at 34 K. In contrast, the high-field value of the differential Hall coefficient decreases by more than a factor of two over the same temperature interval. 
 
The principal observation of this work is that the high-field differential Hall coefficient saturates below 15 K to a temperature- and magnetic field-independent value. We define this high-field saturation value of differential Hall coefficient as $r_{\,\H}^*$, as shown in Figure 3A. The saturation of $r_{\,\H}$ at high magnetic fields and low temperatures is also clearly evident in the raw Hall resistivity shown in Figures 2A,B.
To understand the significance of the saturation temperature of 15 K at 60 T we compare Hall resistivity to the longitudinal resistivity. Figure 3B shows the temperature dependence of resistivity (replotted from Figure 1) and of the differential magnetoresistance $ β ≡  dρ_{xx} / dB $. At  60~T, the differential magnetoresistance  saturates below 15 K, marking the onset of the high-field regime in the longitudinal resistivity, $ ρ_{xx}  \sim  \text{max} \curly{ α^* T , β^* B } $, where the resistivity is $B-$linear. 
\footnote{The ratio between the crossover temperature, 15~K, and magnetic field, 60~T, approximately corresponds to the ratio between the effective thermal and magnetic energy scales as determined by the slopes of the $T-$linear resistivity $α^* = dρ_{xx} / dT = 1.07  \, μΩ\text{cm}/K$  at high temperatures, $ k_{\B} T \gg μ_{\B} B$  (Figures 2A,3D) and $B-$linear resistivity $β^* = dρ_{xx} / dB  = 0.32 \, μΩ\text{cm}/T$ at high magnetic fields, $μ_{\B} B \gg k_{\B} $ (Figure 3A,B). The field and temperature of the onset of the high-field regime in the longitudinal resistivity follows a simple linear relation: the energy scales associated with the crossover at 15 K and 60 T evolve in such a way that at higher temperatures the high-field regime sets in at correspondingly higher magnetic fields. \cite{Giraldo-Gallo2018} }

 The saturation of the differential Hall coefficient at high magnetic fields marks the onset of the high-field regime—the same high-field regime that is observed in the longitudinal resistivity. 
In a conventional Fermi liquid, $ρ_{xy}$ in the high field regime has a slope set by the carrier density and extrapolates to zero as a function of the magnetic field. In contrast, our $ρ_{xy}$ has a finite offset in the high field regime, which we define to be $q_{\,\H}$. This offset is zero at low temperature and increases linearly up to a temperature where 60 tesla is insufficient for $ρ_{xy}$ to be in the high-field regime (Figure 3C). The $T-$linear behavior of $ q_{\,\H}$ indicates that the Hall coefficient in the strange metal state loses its direct relation to charge carrier density. Instead, such temperature dependence of the offset $q_{\,\H}$ indicates  this scale-invariant behavior of the Hall resistivity. \footnote{ The functional form of the Hall resistivity in the high-field regime is $ρ_{xy} = q_{\,\H} + r_{\,\H}^* B$ where  $q_{\,\H}$ vanishes at zero temperature and increases approximately linearly with increasing temperature, $q_{\,\H}\propto T$  (Figure 3A).  } Such scale invariant behavior of the Hall resistivity might be expected in a straightforward extrapolation of the strange metal phenomenology to include the Hall resistivity. However, unlike the longitudinal resistivity, which fits naturally into the strange metal phenomenology, the high-field regime for Hall resistivity is defined not only by field-temperature competition but also by competition with the scale $Λ$ that defines the low-field behavior. One possible reason for the energy scale $Λ$ appearing in the Hall resistivity is that Hall resistivity is constrained by Onsager reciprocity to be zero at zero magnetic field, even in strange metals.\cite{Onsager1931} This does not explain why the Hall coefficient has a strong temperature dependence in the cuprates at high temperatures, i.e., in the low-field regime,  but it gives a rationale for its non-scale-invariant form in this regime.
 
The Hall resistivity at high magnetic fields has been used previously to infer how the carrier density changes as a function of hole doping.\cite{Badoux2016,Ayres2021} In our data, the saturation value of the differential Hall coefficient in the high-field regime is $r_{\,\H}^* = 0.050 \, μΩ\text{cm/T} = 80 \, \text{Å}^3 / e $  (Figure 3A). For the composition of our sample–1.2 holes per planar copper–this is the same value that one expects of the Hall coefficient $R_{\,\H}$ of a single-band Fermi liquid metal. However, in a conventional metal, $R_{\,\H}$ gives the carrier density only in the high-field regime once $ω_c τ \gg 1$ where there is no temperature-dependence to $R_{\,\H}$ and therefore, Hall resistivity has zero offset. By contrast, the Hall resistivity in the high-field regime in our measurement shows the $T-$linear offset $q_{\,\H}$ that is not expected from conventional Fermi liquid phenomenology. Therefore, the crossover to the high-field regime in our data is distinct from the crossover in conventional metals. Instead, our measurement suggests that the crossover into the high-field regime of the Hall resistivity is driven directly by the energy scale competition in the dynamics of relaxation rate. 

If the observed field-temperature competition in the longitudinal resistivity is evidence for a direct effect of the magnetic field on the dynamics of the relaxation rate, then our observation of field-temperature competition in the Hall resistivity suggests that the magnetic field also induces a “skew” character to the relaxation dynamics. This suggests an alternative mechanism for the generation of the Hall voltage in a strange metal: rather than being generated by the Lorentz force acting on quasiparticles, the Hall is generated directly by the field-induced skew character of the scattering rate itself.

{\bf Acknowledgments.}  
The measurements in the 65 T and 100 T pulsed magnet systems were performed at the NHMFL’s Pulsed Field Facility, supported by the NSF through DMR-1644779, the state of Florida and the U.S. Department of Energy. A.S., R.McD., M.-K.C., and F.F.B. acknowledge support from the DOE/BES “Science at 100T” grant. A.S. acknowledges the hospitality of the Aspen Center for Physics, where part of the analysis was performed. Aspen Center for Physics is supported by the NSF grant PHY-1607611. Y.L. acknowledges support from the US Department of Energy through the LANL/LDRD program and the G.T. Seaborg institute. B.J.R. acknowledges funding from the National Science Foundation under grant no. DMR-1752784.


%

	\cleardoublepage
	\begin{figure}[ht!!] \label{fig:1} 
	\includegraphics[width=0.85\textwidth, keepaspectratio]{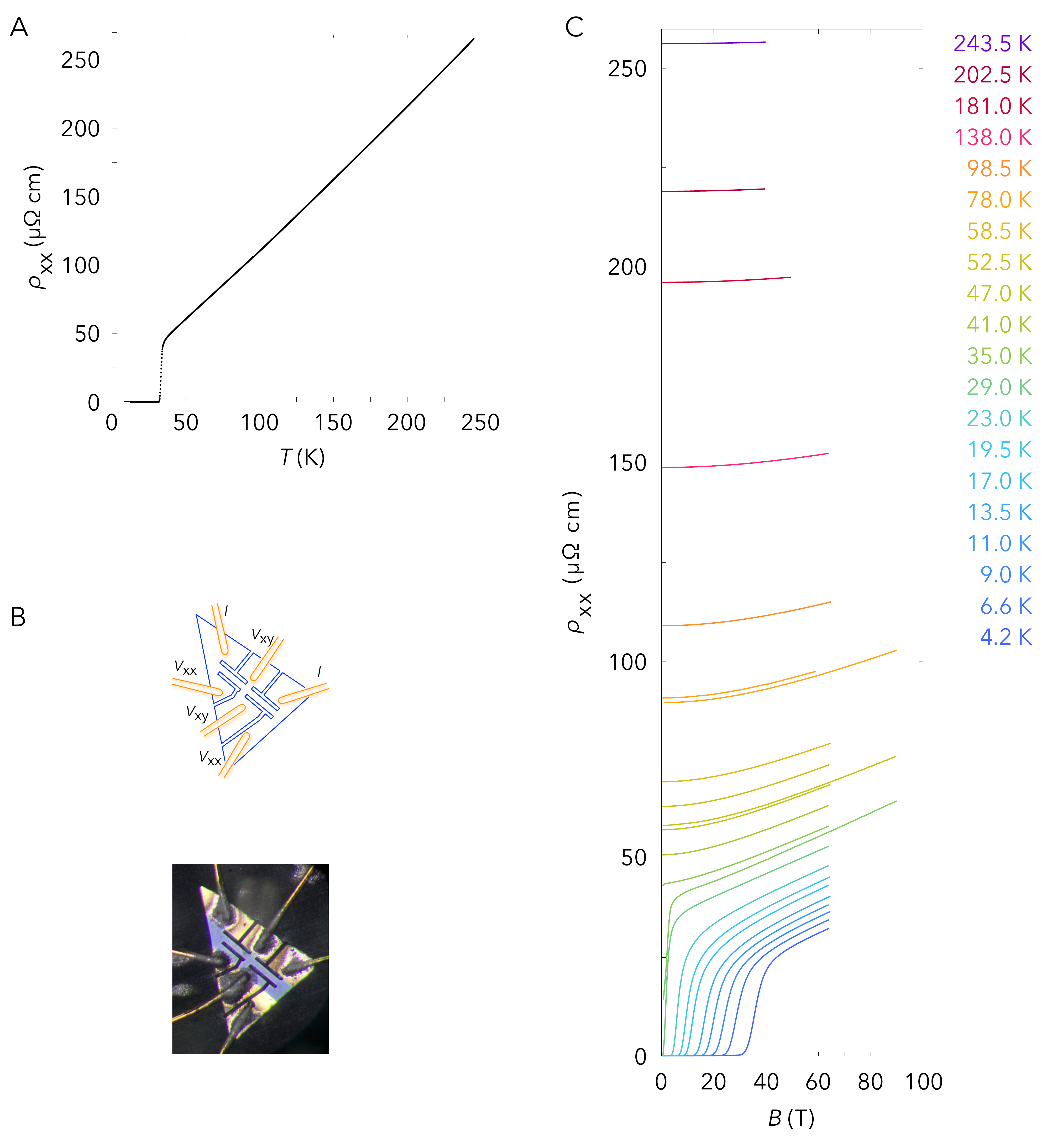} 
	\caption{
{\bf A} Temperature dependence of the longitudinal resistivity $ρ_{xx}$.  
{\bf B} Optical image and schematic of the \LSCO sample used in this measurement. The 350 $μ$-long and 30 $μ$-wide Hall bar was cut out of 12 $μ$-thick \LSCO platelet hand-polished parallel to the copper-oxide plane. 
{\bf C} The longitudinal resistivity $ρ_{xx}$.  The data taken in the 100 T magnet system is at 32.5 K, 50.5 K, 78.5 K. 
	} 
\end{figure} 

	\cleardoublepage
	\begin{figure}[ht!!] \label{fig:2} 
	\includegraphics[width=0.85\textwidth, keepaspectratio]{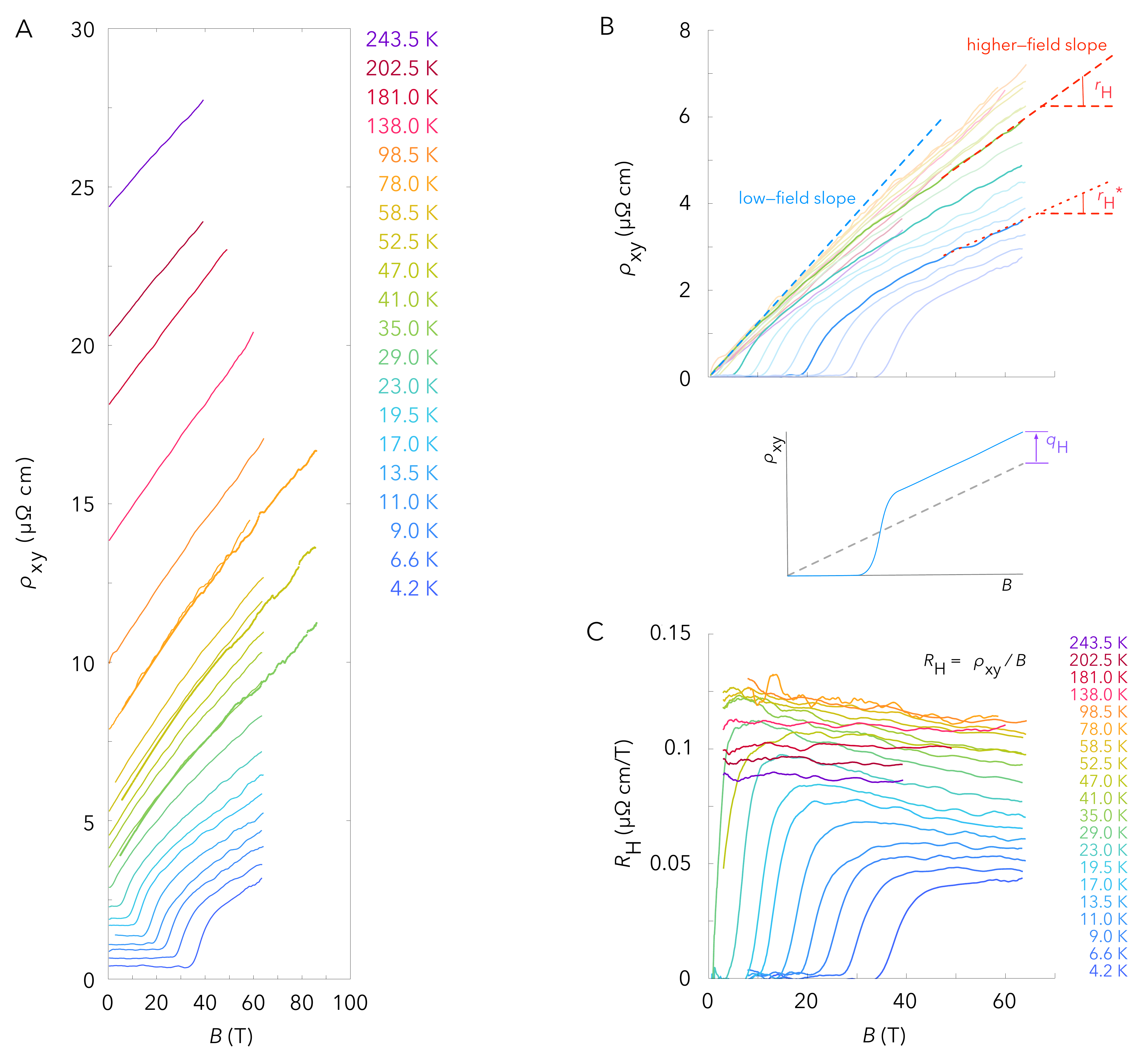} 
	\caption{
Magnetotransport in critically doped \LSCO, $x=0.20$ ($T_c = 34 K$). 
{\bf A} Hall resistivity $ρ_{xy}$ for the same set of temperatures as $ρ_{xx}$. For clarity, each curve is offset vertically in proportion to its temperature,  
$Δρ_{xy}$(μΩ cm) $=$ 0.1 T (K). Each data for $ρ_{xy}(B)$ has been antisymmetrized using two field sweeps with opposite magnetic field directions.  
{\bf B} Magnetic field dependence of the Hall resistivity, replotted from Figure 2A without vertical offsets. For clarity, all field sweeps except for 11 K, 23 K and 35 K are shown in half tone. 
The blue dashed line indicates the field-slope $r_{\,\H} = dρ_{xy} / dB$ at low magnetic fields (where it is coincident with the Hall coefficient, $R_{\,\H} = ρ_{xy} / B$ ) for the field sweep at 35 K, right above the superconducting transition. The red dashed line indicates the high-field slope, $r_{\,\H} = dρ_{xy} / dB$, at 60T, for the same temperature 35 K. As temperature decreases, the value of $r_{\,\H}$  at 60 T decrease monotonically and saturates below 15 K to the value denoted as $r_{\,\H}^*$, indicated by the dotted red line for the field sweep at 11 K. The saturation of $r_{\,\H}$ at high magnetic fields and low temperatures marks an onset of the high-field regime where the Hall resistivity is linear-in-field, $ρ_{xy} = q_{\,\H}+ r_{\,\H}^* B$, where $q_{\,\H}$ is a finite temperature-dependent offset. The high-field offset $ q_{\,\H} $ vanishes at zero temperature. 
{\bf C}  Magnetic field dependence of the Hall coefficient, $R_H = ρ_{xy} / B $. The monotonic decrease of the Hall coefficient at high magnetic fields is a consequence of the non-Fermi liquid behavior of the Hall resistivity at high magnetic fields. 
	} 
	\end{figure} 

\cleardoublepage
\linespread{0.8}
	\begin{figure}[ht!!] \label{fig:3}	
\includegraphics[width=0.7\textwidth, keepaspectratio]{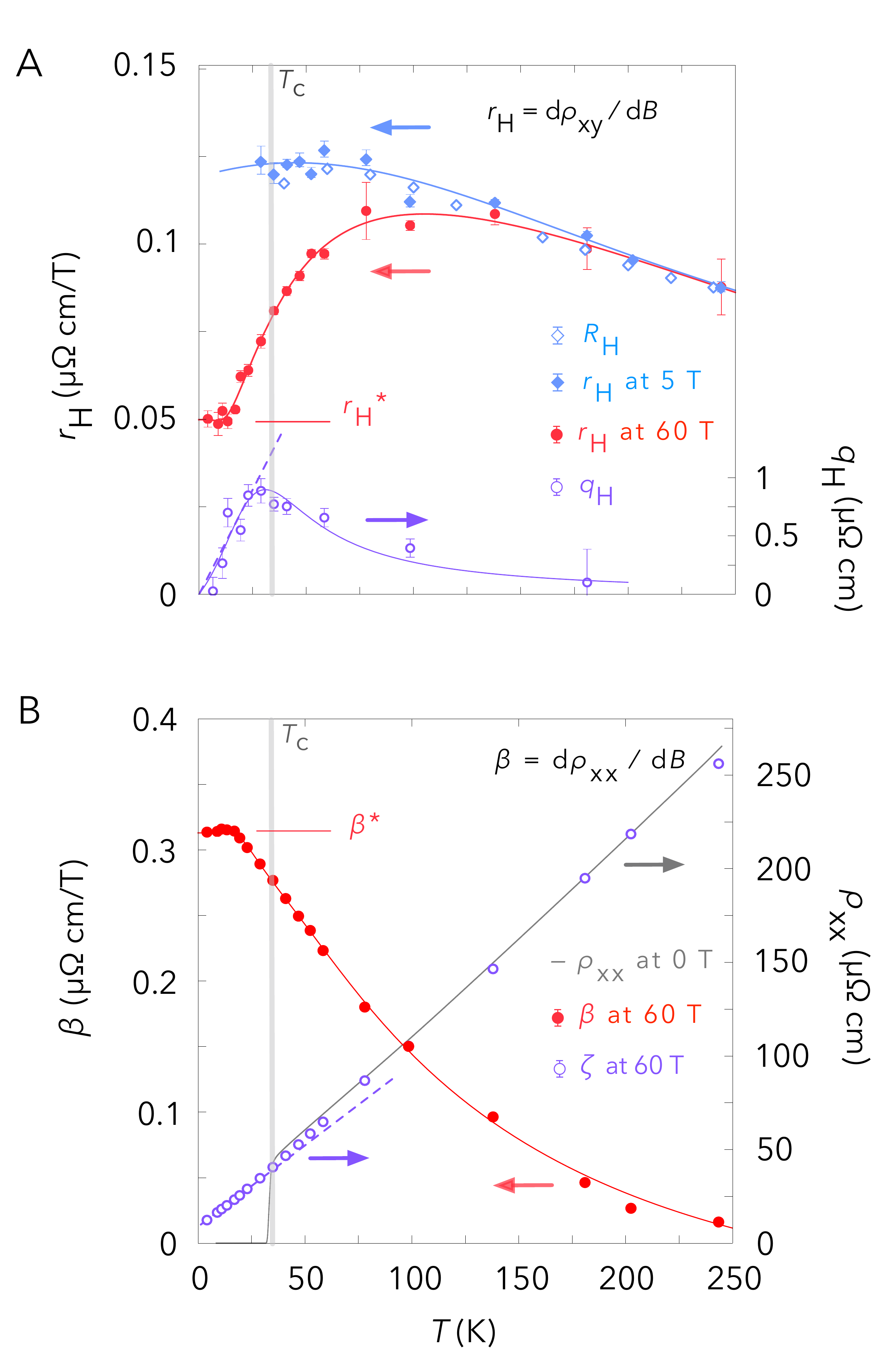} 
\caption{
{\bf A} Temperature dependence of the differential Hall coefficient, $r_{\,\H} = dρ_{xy} / dB$ at 60 T ( filled red circles) and at zero magnetic field (filled blue diamonds). Blue open diamonds indicate measurements of Hall coefficient $R_{\,\H}$ on the same sample made in a superconducting magnet. Open purple circles indicate the temperature dependence of the high-field offset $q_{\,\H}(T)$, as defined for the data at 60 T in Figure 2A. Below the saturation temperature of 15 K, the value of $q_{\,\H}(T)$ is approximately linear in temperature, as indicated by the dashed purple line. All solid lines are guides to the eye. Grey vertical line at 34 K indicates the superconducting transition temperature at zero magnetic field. 
{\bf B} Temperature dependence of the differential magnetoresistance, $β = dρ_{xx} / dB$, at 60 T (red filled circles). The black solid line indicates the zero-field resistivity from Figure 1A. Grey vertical line at 34 K indicates the zero-field superconducting transition temperature. The open purple circles indicate the high-field offset $ζ$ in the linear-field behavior of  $ρ_{xx}(B)$  at 60 T in Figure 2 A. Below the saturation temperature of 15 K, the value of the offset $ ζ(T) $ (approximately linear in temperature, as indicated by dashed purple line) determines the limiting behavior of the magnetoresistance in the high-field regime, $ρ_{xx} = ζ + β^* B$. 
} 
	\end{figure}
	
\end{document}